\documentclass[pre,twocolumn,twoside,showpacs,superscriptaddress]{revtex4-1}
\usepackage{graphics,amssymb,amsmath,epsfig,color}
\usepackage{rotating,multirow}
\epsfclipon

\begin{document}
\title{Costly bilingualism model in a population with one zealot}
\author{Hyunsuk Hong}
\email{hhong@jbnu.ac.kr}
\affiliation{Department of Physics and Research Institute of Physics and
Chemistry, Chonbuk National University, Jeonju 561-756, Korea}
\author{Seung-Woo Son}
\email{sonswoo@hanyang.ac.kr}
\affiliation{Department of Applied Physics, Hanyang University,
Ansan 426-791, Korea}
\date{\today}
\begin{abstract}
We consider a {\it costly bilingualism model} in which one can take two strategies
in parallel. We investigate how a single zealot
triggers the cascading behavior and how the compatibility of the two strategies affects when interacting patterns change. First, the role of the interaction range on the cascading is studied
by increasing the range from local to global.
We find that people sometimes do not favor to take the superior strategy even though its payoff is higher than that of the inferior one. This is found to be caused by the local interactions rather than the global ones. Applying this model to social networks,
we find 
that the location of the zealot is also important for larger cascading in heterogeneous networks.
\end{abstract}
\maketitle

\section{Introduction}

One of the fundamental issues in social science is to understand how new strategies, technologies, and ideologies spread and diffuse through population~\cite{ref:idea_strategy_spread,Marvel2012}.
One possible mechanism that may explain this phenomenon is {\it{bilingualism}}, where people can adopt two traits -- such as languages, technologies, and ideas -- in parallel. Many of the related researches are performed also in physics as well as economics and mathematical sociology~\cite{Marvel2012,ref:JonKleinberg,ref:JonKleinberg_1}.

In the recently studied bilingualism models by Kleinberg et al., a population with two early adopters has been considered and how the compatibility of bilingualism influences the cascading behavior has been examined~\cite{ref:JonKleinberg,ref:JonKleinberg_1}. Motivated by these studies, we investigate the
population with single zealot since we realized even one
zealot can trigger off the cascades. In particular, we focus on how the interaction range
influences the cascading behavior, and explore the system by systematically changing the range from local to global. Moreover, in order to see how the compatibility affect competition in the real world, we also
apply the model to the real social networks, and investigate its cascading behavior.

This paper is organized as follows: a costly bilingualism model is introduced in Sec.~\ref{model}, then we reports the analytic and numerical results on one dimensional ring with an increasing interaction range in Sec.~\ref{results1} and those of the globally coupled case in Sec.~\ref{results2}. In Sec.~\ref{application}, an application to social networks are displayed. Finally we summarize and discuss our results in Sec.~\ref{summary}.

\section{Costly Bilingualism Model}
\label{model}

\begin{table}[b]
\begin{tabular}{|c|c|c|c|c|}
    \hline
    \multicolumn{2}{|c|}{}&\multicolumn{3}{|c|}{neighbors' strategy}\\
    \cline{3-5}
    \multicolumn{2}{|c|}{}& $A$ & $B$ & $AB$ \\
    \hline
    \multirow{3}{*}{\begin{sideways}~strategy~\end{sideways}}
                        & $A$ & $a$ & $0$ & $a$ \\
    \cline{2-5}
                         & $B$ & $0$ & $b$ & $b$ \\
    \cline{2-5}
                         & ~$AB$~ &~~~~~~$a-c$~~~~~~&~~~~~~$b-c$~~~~~~& max$(a,b)-c$ \\

    \hline
\end{tabular}
\caption{Payoff matrix of the costly bilingualism model.  Here
max$(a,~b)$ represents the larger of $a$ and $b$, and $-c$
in the last row denotes the cost of bilingualism.}
\label{table:payoff}
\end{table}

Let us consider a population where individuals can choose
one of the three strategies $A$, $B$, and $AB$.  $A$ and $B$ can be
regarded as two languages, and $AB$ represents
bilingualism. Learning multiple languages or technologies usually requires lots of time and resources and thus we assume that the bilingual strategy is {\it{costly}}. The costly bilingualism model is played as follows~\cite{ref:JonKleinberg, ref:JonKleinberg_1}:
If two individuals with the same strategy $A$ (or $B$) interact, they get the same payoff $a$ ($b$).
If two with different strategies -- one with $A$ and the other with $B$ -- interact, no one gets any payoff.
When a bilingual ($AB$) interacts with a monolingual, each gets the monolingual's payoff. The bilingual
individual pay the cost $c$ of adopting bilingualism. When two bilingual individuals interact with each other they get the payoff of max$(a,b)$ while paying the same cost $c$.
The payoff matrix is shown in Table~\ref{table:payoff}.

At each time step, each individual updates its strategy choosing one among three strategies to maximize its payoff with probability 1 with no transition cost. While updating one's strategy, we assume there is no change on his/her neighborhood. In this model, a direct update between `monolingual' strategies such as change from $A$ to $B$ are allowed, differently from other model~\cite{Marvel2012} where all shifts between monolingualism have to pass through the `bilingual' strategy $AB$. This cascading update continues until no one wants to change anymore. Note that we do not concern the details of how one can change his/her strategy and its transition cost. It can be an imitation or adopting an already known-technology freely, to maximize players' utility function.

We introduce a {\it{zealot}}, who does not change his/her strategy from the strategy $A$ at all, in the population of $B$. Then we study the $A$'s cascading behavior for various values of $a$ and $c$, under a fixed value of $b$, where $b$ is set to 1, and $a$ is assumed to be larger than $b$. When the number of interacting neighbors is more than one, we rescale the value of $a$ and $b$ by its number of neighbors to keep a balance with the cost $c$. For example, when one interacts with three neighbors using $A$, the individual gets the payoff $a/3$ per neighbor.

\section{Results}
\subsection{Local Interactions}
\label{results1}
We first consider one dimensional ring with an interaction range $\ell$, where each individual interacts with $2\ell$ neighbors. $\ell=1$ means the conventional one-dimensional lattice that
each individual has two nearest neighbors, and $\ell=2$ represents the
system where each one interacts with four neighbors (nearest neighbors and the next
nearest neighbors). We study how the cascading behavior changes as we vary the interaction range from the local $(\ell=1)$ to the global $(2\ell=N-1)$.

\begin{figure}[b]
\centering
\epsfxsize=0.475\linewidth \epsfbox{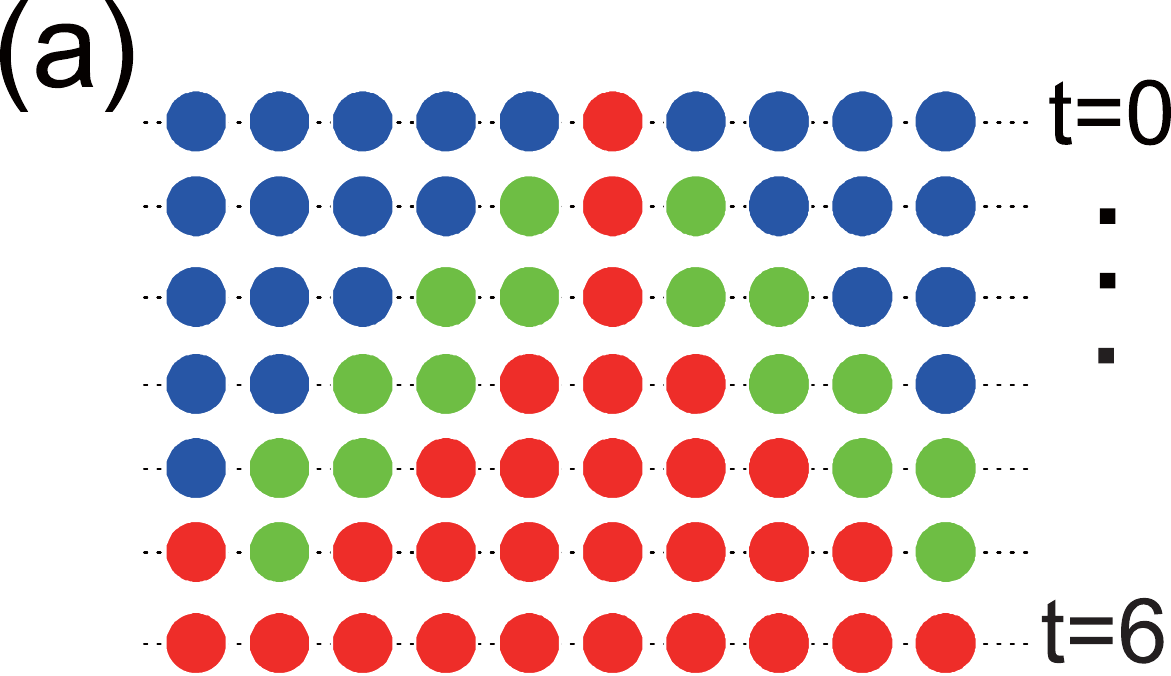}
\epsfxsize=0.475\linewidth \epsfbox{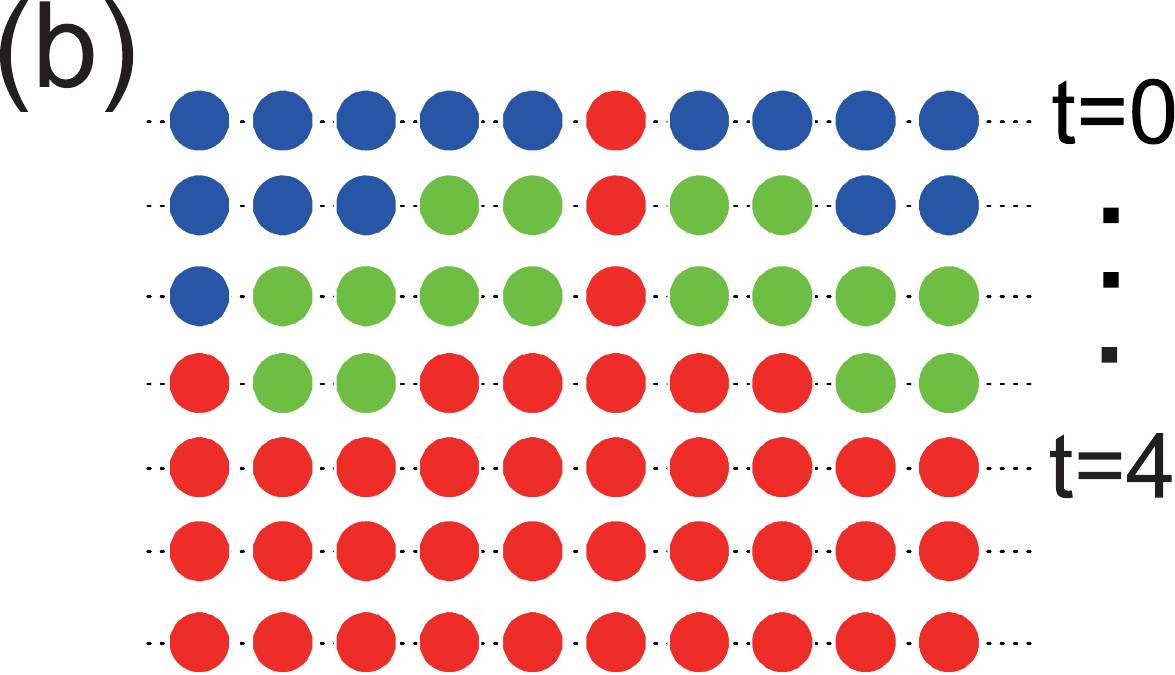}
\caption{(Color online)
Schematic plots of strategy updates for the interaction range
(a) $\ell=1$, and (b) $\ell=2$ with the payoff $a=2$, $b=1$,
and the cost $c=0.5$. Red circles represent the strategy $A$ (the
one at $t=0$ is the zealot who maintains the strategy $A$), the blue ones $B$, and the
green $AB$, respectively.
}
\label{fig:dynamics}\end{figure}

Figure~\ref{fig:dynamics}(a) and (b) show how the
individuals updates their strategy for $\ell=1$, and for $\ell=2$ with the payoff $a=2$, $b=1$, and the cost $c=0.5$~\cite{note1}. For $\ell=1$, two nearest neighbors with $B$ 
next to the $A$ zealot (colored in red at $t=0$ in Fig.~\ref{fig:dynamics}(a))
first adopt $AB$ at $t=1$. The next nearest neighbors then adopt $AB$,
and the former two $AB$s change into $A$. At $t=6$, all takes $A$ and the cascade completes.
The strategy $AB$ naturally emerges even though we start with no $AB$.
The person who choose the strategy $B$ usually adopts $AB$ first, and then
changes into $A$, but directly adopt $A$ when the cost $c$ is too high.
For $\ell=2$, on the other hand, the four nearest neighbors next to the
zealot first adopt $AB$ at $t=1$.
At $t=2$, the other four next nearest neighbors adopt $AB$, and
the four formerly adopting $AB$s change into $A$ at $t=3$. The cascade completes at $t=4$, as shown in Fig.~\ref{fig:dynamics}(b).
The duration of cascades decreases as the interaction range $\ell$ increases.

\begin{figure}[t]
\centering
\epsfxsize=0.485\linewidth \epsfbox{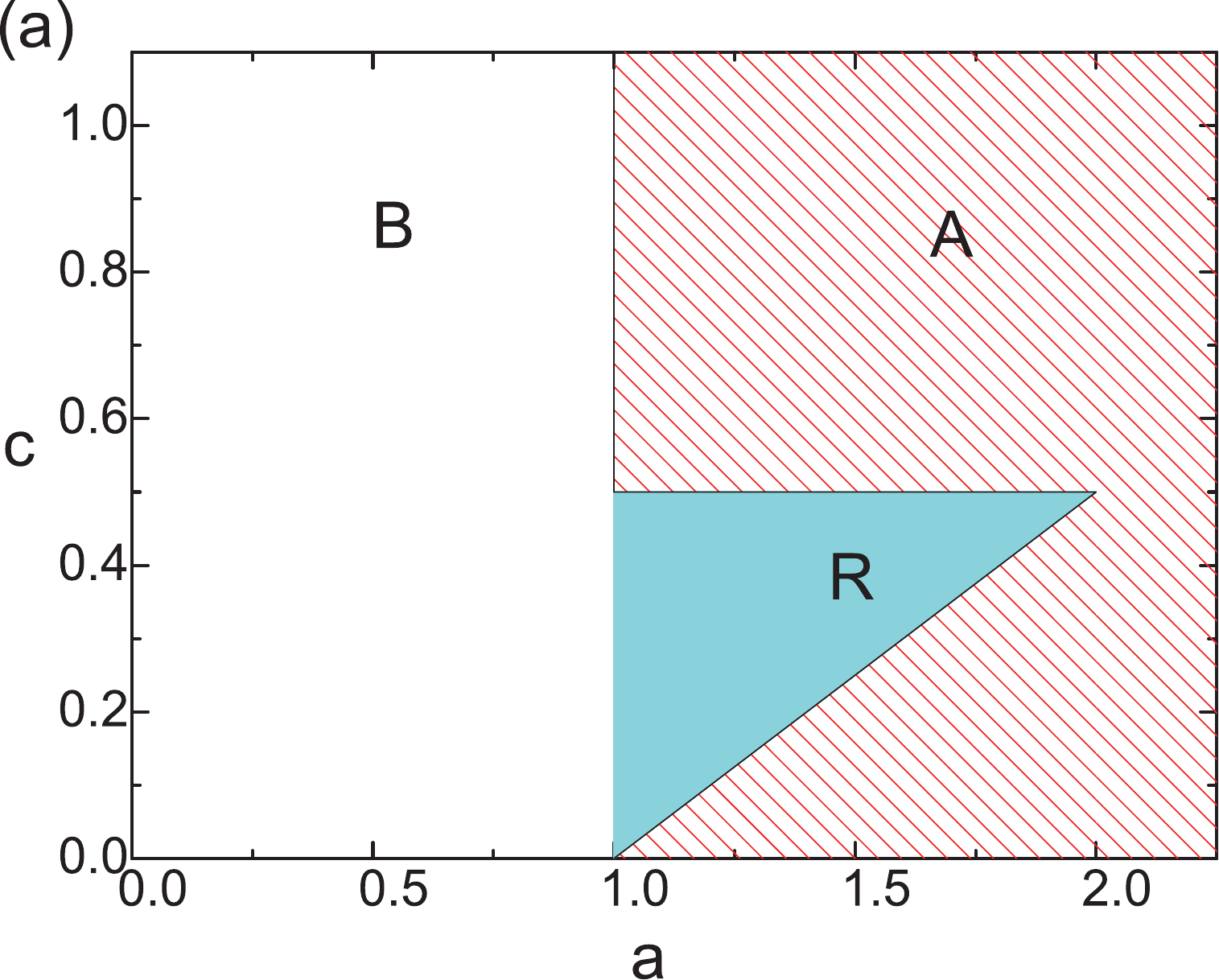}
\epsfxsize=0.485\linewidth \epsfbox{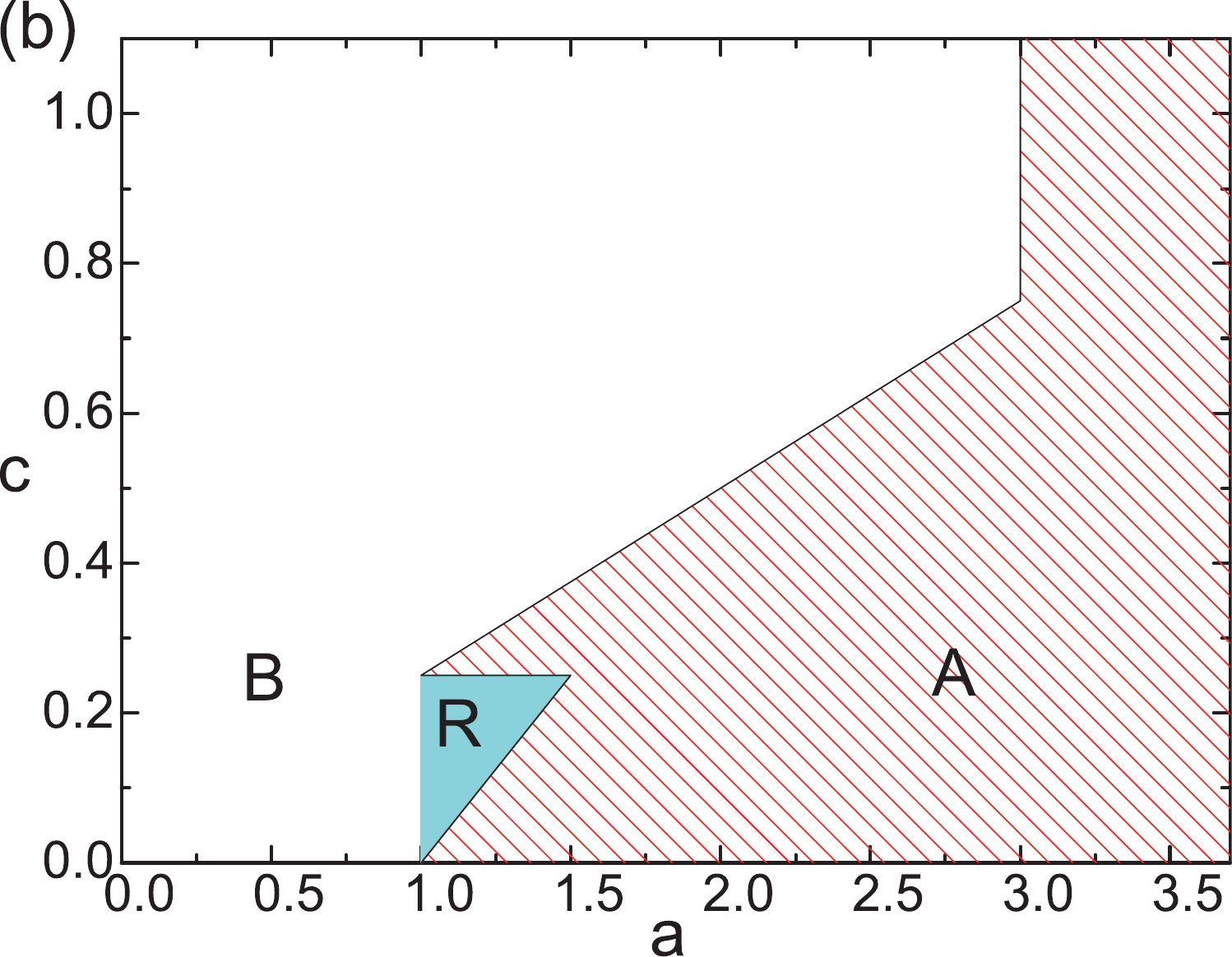}
\epsfxsize=0.485\linewidth \epsfbox{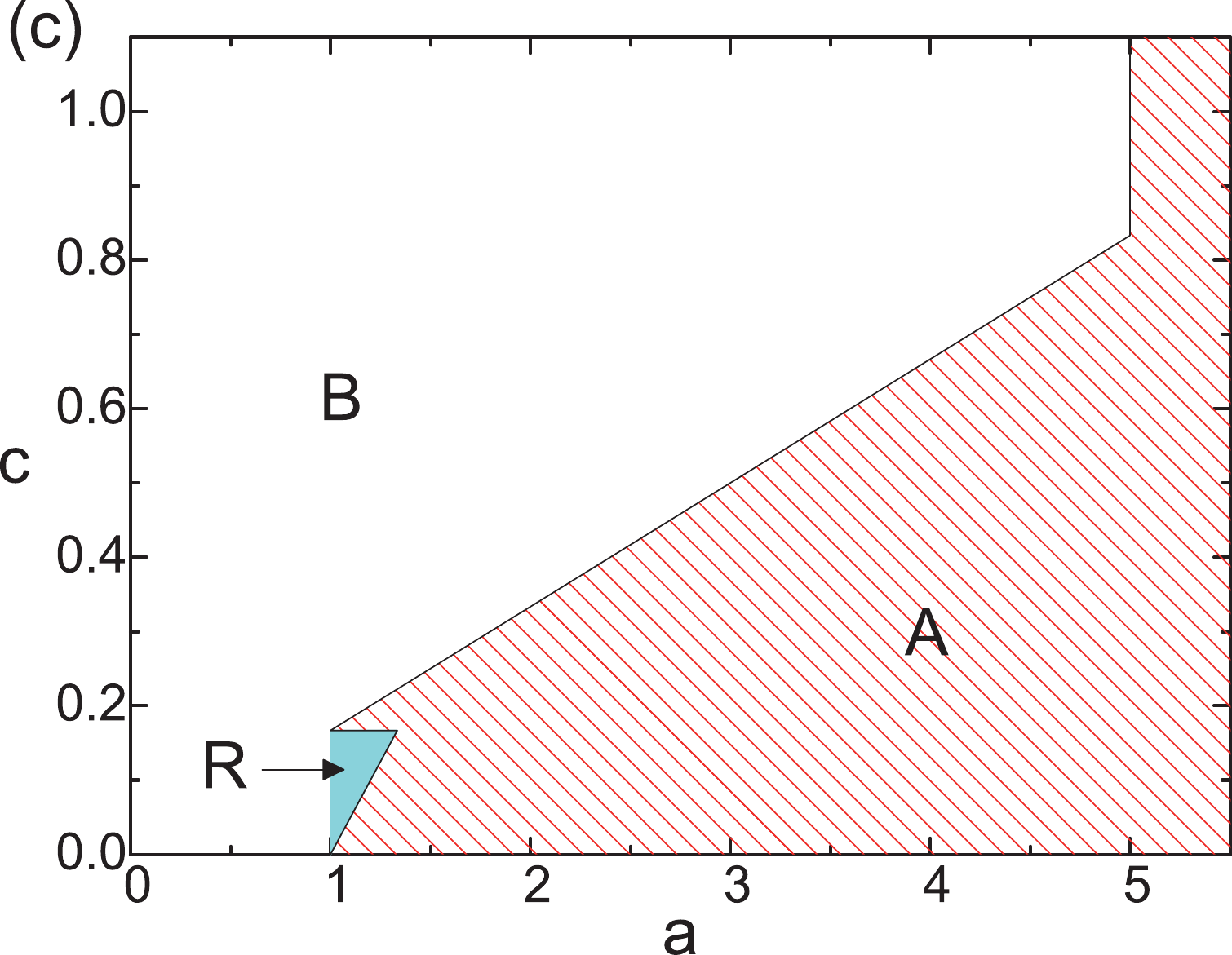}
\epsfxsize=0.485\linewidth \epsfbox{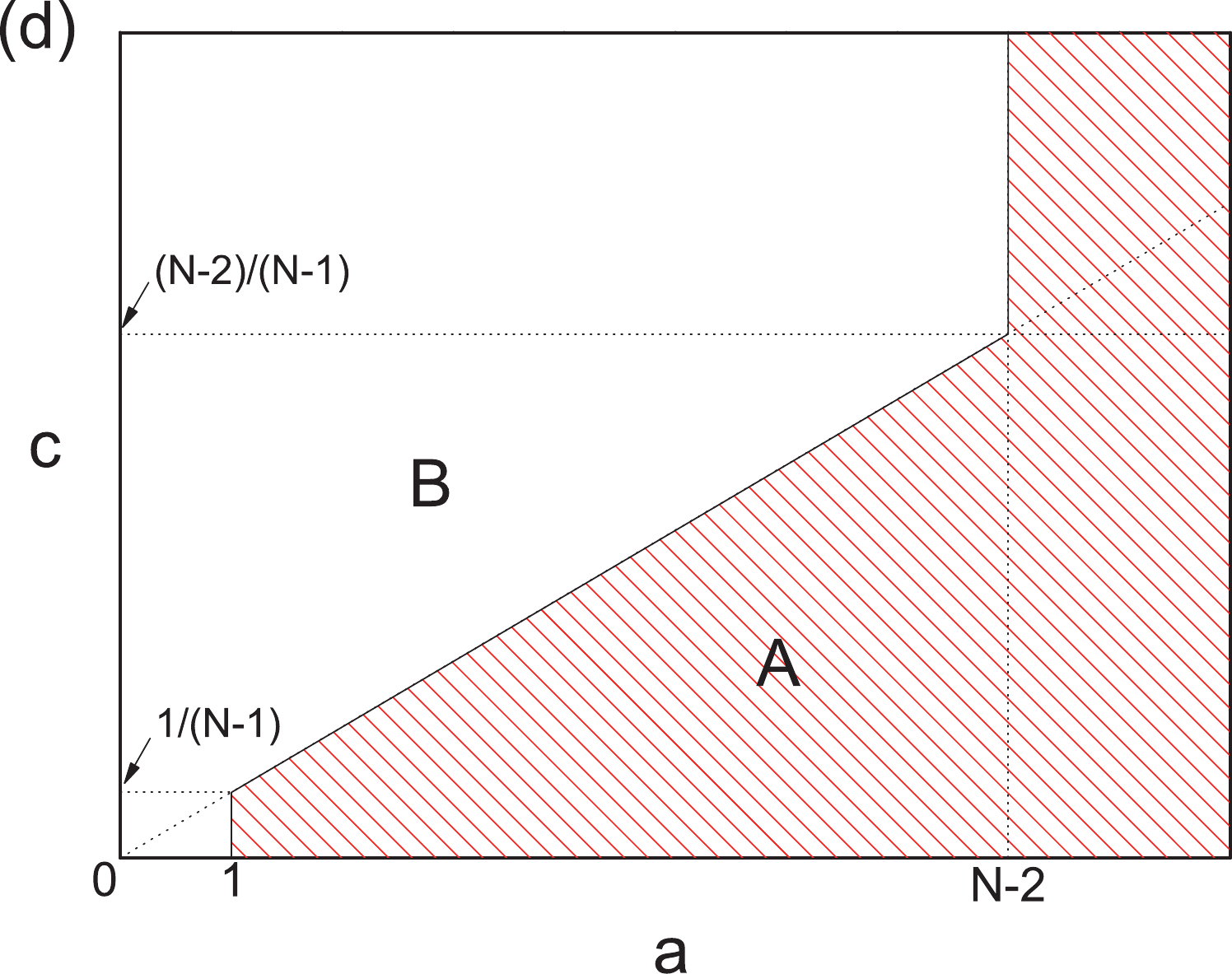}
\caption{(Color online) Phase diagram in the plane of $a$ and $c$ with $b=1$
for (a) $\ell=1$, (b) $\ell=2$, (c) $\ell=3$, and (d) globally-coupled version of
the costly bilingualism model, respectively.
For the phases labeled ${\bold{A}}$, ${\bold{B}}$, and ${\bold{R}}$
[see the text].
}
\label{fig:phd}\end{figure}

We investigate the cascading behavior at various values of
the interaction range, payoff, and cost.
Figure~\ref{fig:phd} shows the phase diagram in the plane of
$a$ and $c$ for various interaction ranges $\ell$ from $\ell=1$ to
$\ell=3$ [Fig.~\ref{fig:phd}(a)-(c)], and the globally-coupled version [Fig.~\ref{fig:phd}(d)], respectively.
In particular, we note that the phase diagram shown in
Fig.~\ref{fig:phd}(a)-(c) is divided
into three regions: Two phases labeled (in bold) ${\bold A}$ and ${\bold B}$,
where ${\bold A}$ represents the phase that all individuals in the
population choose the strategy $A$ in the long time, and ${\bold B}$
denotes the phase that all (except the one $A$-zealot) takes the
strategy $B$.
The other one is the phase ${\bold{R}}$, where the strategy $AB$ survives at the boundary and
blocks the $A$'s spread through $B$, making the cascade of $A$ impossible.
This ${\bold{R}}$ phase appears in the carved region
of the phase diagram in Fig.~\ref{fig:phd}(a)-(c).
Interestingly, people in this phase does not favor to take the
strategy $A$ even though the payoff $a$ from taking $A$ is larger
than the payoff $b$ from the strategy $B$.
We note that the phase ${\bold{R}}$ consists of $2\ell$-wide $AB$s located at
the next to the $A$ zealot and all $B$s for the others, so the number of
$AB$s increases as the interaction range $\ell$ increases.
In the ${\bold{R}}$ phase, the $AB$s shield the
$A$-zealot, which makes a $AB$-{\it buffering} zone, prohibiting
the cascade.

It is noteworthy to point out that {\it reentrant transition}
occurs in the phase ${\bold{R}}$, e.g., for $\ell=1$ the population shows the
$A$-cascade for $a=1.5$ and $c=0.1$, but if we increase the cost up to
$c=0.3$, the $AB$s next to the zealot shield the $A$, blocking the
$A$'s spread, which impedes the $A$'s cascade.
However, if we further increase the value of $c$ up to $c=0.6$ the
population reaches the cascade again~[see Fig.~\ref{fig:phd}(a)].
We also find that the area of this reentrant-transition zone shrinks as $\frac{b^2}{4{\ell}^2}$ as shown in
Fig.~\ref{fig:phd}(a)-(c) [see later], and the zone eventually
disappears in the globally-coupled system as shown in
Fig.~\ref{fig:phd}(d).  This implies that the phase ${\bold{R}}$
is caused by the local interaction not by the long-range (global) one.
We note that the occurrence of the phase ${\bold{R}}$ has also been
reported in Ref.~\cite{ref:JonKleinberg}, where the authors
studied for $\ell=1$ case with the two early adapters (zealots).
Our study, on the other hand, exhibits that even only one zealot
can trigger off the cascading, furthermore
we find that the phase ${\bold{R}}$ strongly depends on the interaction
range $\ell$.

\subsection{Globally Coupled Case}
\label{results2}

We now consider the globally-coupled version of the model, i.e.,
the complete graph with the population size $N$, where each individual interacts
with all the others, and analyze the behaviors of the system.
The phase boundary between the phases ${\bold{A}}$ and ${\bold{B}}$ in
Fig.~\ref{fig:phd}~(d) can be obtained by analyzing the break-even point
of the payoff.
The payoff of the $i$th individual,
from choosing the strategy $A/B/AB$,
is given by
\begin{eqnarray}
\label{eq:payoff_MF_0}
& &p^{i}_A = \tilde{a}(N^{i}_A+N^{i}_{AB}),~~
p^{i}_B = \tilde{b}(N^{i}_B+N^{i}_{AB}), \\
& &p^{i}_{AB} = \tilde{a}(N^{i}_A+ N^{i}_{AB})+\tilde{b}N^{i}_{B}-c \nonumber
\end{eqnarray}
for $i=1,\cdots,N$, where $N^{i}_A$, $N^{i}_B$, and $N^{i}_{AB}$
represent the number of the interacting neighbors of the $i$th individual,
who plays $A$, $B$, and $AB$, such that
$N^{i}_A+N^{i}_B+N^{i}_{AB}=k^{i}$, where $k^{i}$ is the
number of neighbors (degree) of the $i$-th individual.
And $\tilde{a}=a/k^{i}$ and $\tilde{b}=b/k^{i}$ denote the
rescaled payoff for the balance with the cost $c$.
For the complete graph,
all people has the same number of neighbors ($k^{i}=N-1$) with no difference, which allows us to suppress the
site index $i$ in the Eq.~(\ref{eq:payoff_MF_0}), which reads $p_A = a \left(1- \frac{N_B}{N-1} \right)$, $p_B = b \left(1- \frac{N_A}{N-1} \right)$, and $p_{AB} = a-c-(a-b) \frac{N_B}{N-1}$.

We now start with the initial condition that there is only
one $A$-zealot ($N_A=1$) and the others are $B$'s ($N_B=N-1$).
To choose the strategy $A$, the payoff $p_A$ should be larger than the other
ones $p_B$ and $p_{AB}$.  Similarly, to take the strategy $B$ the payoff $p_B$
should be larger than $p_A$ and $p_{AB}$, and to take the strategy $AB$
the payoff $p_{AB}$ should be larger than the others.
From these conditions, we find that the strategy $A$ is chosen for
\begin{equation}
\label{eq:forA}
a > b (N-2)~ {\rm and}~ c > b \frac{N-2}{N-1},
\end{equation}
the $B$ is chosen for
\begin{equation}
\label{eq:forB}
a < b(N-2)~ {\rm and}~ c > \frac{a}{N-1},
\end{equation}
and the $AB$ is chosen for
\begin{equation}
\label{eq:forAB}
c < b \frac{N-2}{N-1}~ {\rm and}~ c < \frac{a}{N-1}.
\end{equation}
We find that Eq.~(\ref{eq:forB}) with the condition $a>b$ (initially assumed)
determines the phase boundary of the phase ${\bold{B}}$,
as shown in Fig.~\ref{fig:phd}(d).
On the other hand, Eq.~(\ref{eq:forA}) and (\ref{eq:forAB}) decide
the boundary of the parameter region where the strategies $A$ and $AB$ are
chosen, respectively.
Let us suppose that we are now in the parameter region where $AB$ is chosen.
For a given value of $a$ and $c$ in this region, the individuals
first choose $AB$, the system then consists of one $A$-zealot and $AB$s
for the remains, i.e., $N_A=1$, $N_B=0$, and $N_{AB}=N-1$, since all
$B$s except the $A$-zealot turn into $AB$ due to the
``all-to-all'' coupling in the complete graph.
We find that all $AB$s take the strategy $A$ next time since the payoff
obtained from taking $A$ is the largest one, which makes
all $AB$s in this region turn into $A$s
and they remain ever since, which yields the phase boundary
shown in Fig.~\ref{fig:phd}(d).

The phase boundary for the system with local interaction in
Fig.~\ref{fig:phd}(a)-(c) can be also obtained from the analysis of
break-even point of the payoffs of $A$, $B$, and $AB$, similarly
to the globally-coupled case.
However, the density-level description of the globally-coupled system is impossible, instead the node(site)-level one is available, i.e, we should
decide the strategy of each node one by one, considering all
available situation.
Substituting $k^{i}=2\ell$ and the rescaled payoff
$\tilde{a} = \frac{a}{2\ell}$ and $\tilde{b} = \frac{b}{2\ell}$ into
Eq.~(\ref{eq:payoff_MF_0}), the payoff of each individual is obtained, and
the same analysis about the break-even point of the payoffs leads us to have
the phase boundary as shown in Fig.~\ref{fig:phd}(a)-(c), where the
boundary has been also confirmed numerically.
We find that the three points which consist of the triangular region of the phase ${\bold{R}}$
located at the carved zone are given by
the three points $(b,~0)$, $(b,~\frac{b}{2\ell})$, and $(b+\frac{b}{\ell},~\frac{b}{2\ell})$.
Accordingly, the size of the phase ${\bold{R}}$-region is given by
$\frac{b^2}{4{\ell}^2}$, and it vanishes in the globally-coupled system.
And the kinked corner point on the side of large $a$ and $c$ for $\ell=2$ and $3$
is found to be given by $(b(2\ell-1),~b-\frac{b}{2\ell})$.

\section{Application to Real Social Networks}
\label{application}

\begin{figure*} 
\epsfxsize=0.9\linewidth \epsfbox{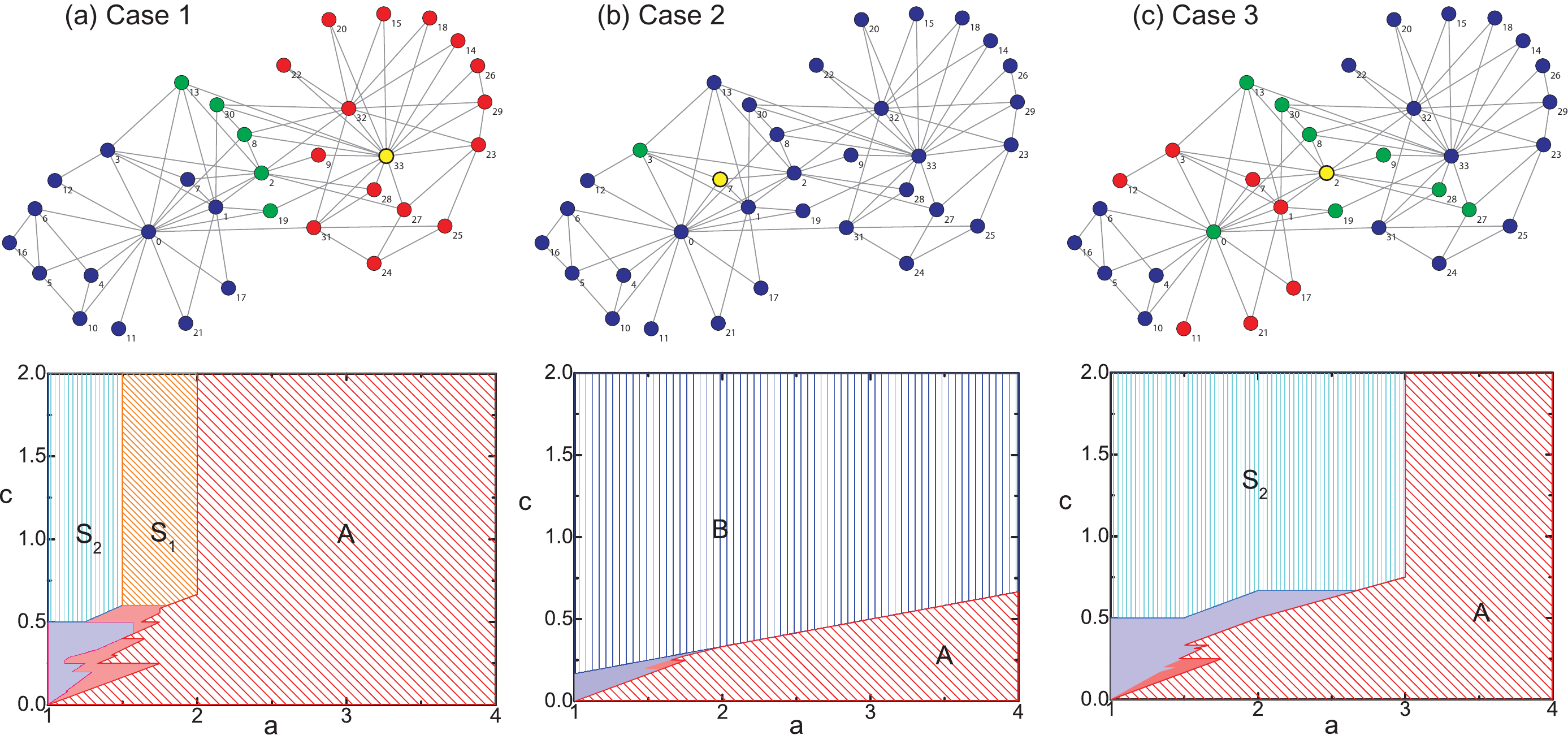}
\caption{(Color online) Final configurations for the costly bilingualism model on
the Zachary's karate club network for $a=1.4$, $b=1$, and $c=0.2$ (top) and phase diagrams for each case (bottom).
Three representative cases are shown
when the zealot (colored in yellow) is located (a) at the node $33$, (b) at $7$,
and (c) at $2$, respectively.
The different colors represent different phases in the phase diagrams [see the text]:
The phase ${\bold{A}}$ is represented by the red hatched lines;
The phase ${\bold{B}}$ by the blue vertical lines;
The phase ${\bold{S_{1}}}$ by the orange fine shaded lines;
The phase ${\bold{S_{2}}}$ by the sky blue fine vertical lines;
The phase ${\bold{S_{3}}}$ by the filled pink region;
The phase ${\bold{S_{4}}}$ by the filled light blue region,
respectively.}
\label{fig:karate}
\end{figure*}

A natural question we can ask now is this: {\it does the reentrant transition
occur in a real social system, too?}  How does the compatibility of the
strategies influence the cascading behavior in the real social networks?
To address these questions, we now analyze real social networks and
explore how the compatibility of the two strategies affect the cascading
behavior.
We consider the Zachary's karate club network~\cite{ref:karate}. The network is known as the social network of
friendships between 34 members of a karate club at a US university in the 1970s.
The network size is just 34, which allows us to do complete
investigation of the cascading behavior depending on the location of the
zealot.  In particular, we examine how the network properties affect the
cascading behavior.

We perform the numerical simulations on the karate club network, where
individuals sequentially update their strategies~\cite{note1}. The sequence of updates is determined according to the expected payoff change; the node that can achieve the largest change in their payoff get the priority
since the large potential payoff change can be considered as a high social pressure from the neighbors.
Again we find that even a single $A$-zealot can trigger the cascades
in the karate club network. In addition, we find that the location of the zealot is crucial.
We consider all 34 possible locations of the zealot, and
explore how the phase diagram changes by the location of the zealot.
We obtain 34 phase diagrams for the different sites of the zealot and find that the phase diagrams can be classified into the three representative ones as shown in Fig.~\ref{fig:karate}.

We find that, in addition to the phases ${\bold{A}}$ and ${\bold{B}}$,
we have more phases named ${\bold{S_1}}$, ${\bold{S_2}}$, ${\bold{S_3}}$,
and ${\bold{S_4}}$, where the phase ${\bold{S_{1}}}$ (${\bold{S_{2}}}$) represents the mixed state of the strategy $A$ and $B$ with no $AB$, while $A$ ($B$) is superior to $B$ ($A$); the phase ${\bold{S_{3}}}$ (${\bold{S_{4}}}$) is the state where $A$, $B$, and $AB$ all coexist, where the $A$ ($B$) is the superior one.
We note that the phases ${\bold{S_3}}$ and ${\bold{S_4}}$ include the
$AB$-buffering zone inside.  The phases are represented
by the different colors as shown in Fig.~\ref{fig:karate}.
Note that there is no other phase except these six.

Interestingly {\it multi-reentrant transition} zone appears as shown in Fig.~\ref{fig:karate},
which is caused by the mixed interaction among the people with a variety of
neighbors.
The network properties
summarized in the Table~\ref{table:networkproperties}
for the node 2, 7, and 33, show that the system easily produces the
cascade when the hub is the zealot, as expected.
This is, however, not enough for achieving the larger cascade.
Additional important conditions are whom the zealot connects and where
he/she is located.  The zealot needs to have small clustering coefficient (CC), and
its neighbors should have small degree. High CC means that
one's neighbors know each other very well and they favor to share a common
strategy, which means that they can convert their strategy at the same time~\cite{Newman2003}.
Therefore, high clustering around one individual can be an obstacle to large cascades.
Furthermore, one's neighbors with small-degree can be easily influenced by the zealot's
opinion since its influence is reciprocally proportional to the
neighbors' degree.
We find that these effects are well observed in the karate
network, as shown in the Table~\ref{table:networkproperties}.

\begin{table}
\begin{tabular}{c|rrrrc}
\hline
node $i$~~&~~~$k^{i}$ &~~~CC~& ~~~$\bar{k}^{i}_{nn}$ &~~~BC~&~~closeness\\
\hline
33&~~17&~~0.11&~~3.8~&~~387.1&~~1.82\\
\hline
7&~~4&~~1.00&~~10.3~&~~66.0&~~2.27\\
\hline
2&~~10&~~0.24&~~6.6~&~~217.7&~~1.79\\
\hline
\end{tabular}
\caption{Network properties. $\bar{k}^{i}_{nn}$ means the average neighbor degree, ``BC'' means
the betweenness centrality, and ``closeness'' represents closeness centrality,
respectively~\cite{Newman2003}.}
\label{table:networkproperties}
\end{table}

On the other hand, the case of the network with strong community
structures can be a different story.
We investigate several target nodes in other social networks:
Les Miserables network~\cite{Knuth1993} ($N=77$),
dolphin network~\cite{Lusseau2003} ($N=62$),
and coauthorship network of network scientists~\cite{Newman2006}
($N=379$, only considering the giant connected component),
based on high $k^i$, small CC, and small
$\bar{k}^{i}_{nn}$.
We find that small CC, small $\bar{k}^{i}_{nn}$,
and even high $k^i$ do not promise large cascade size since cascades often stop after converting several nodes in a community even when they start from the hubs. Therefore, the size of assigned community or the centrality can be
also important factors for the cascades when the network has high
modularity, which requires an in-depth study on the
role of the modularity on the cascading in complex networks.

\section{Summary and Discussion}
\label{summary}

To summarize, we considered costly bilingualism model in a
population with one zealot, and explored how the compatibility
influences on the cascading behavior of one strategy, extending
the interaction range from local to global.
We found that superior strategy does not necessarily propagate.
In the parameter region where this phenomenon occurs, the {\it reentrant} phase transition occurs.
We found that it is caused by the local interaction with one's neighbors rather than the
long-range one. We applied the model to real-world social network and showed how the network properties take effects in the cascades. We have learned the lessons that if the zealot locates at the node with high degree,
this is good for the larger cascade. Furthermore, we showed that the small clustering coefficient and small average neighbors degree enhance the cascade. Finally, we demonstrated that the community structure makes it hard to predict the cascade size. If a network has high modularity, the community structure
should be considered carefully, which is remained for future study.

\begin{acknowledgements}
This work was supported by National Research Foundation of Korea (NRF) Grant
funded by the Ministry of Science, ICT \& Future Planning No. 2012R1A1A2003678 (H.H)
and No. 2012R1A1A1012150 (S.-W.S).
We thank Steven H. Strogatz for suggesting this problem and having useful discussion during the course of this work.
\end{acknowledgements}

\end{document}